\documentclass{pnastwo}
\usepackage{graphicx}
\usepackage{url}
\usepackage{amssymb,amsfonts,amsmath,cite}

\begin{document}

\title{Synchronicity, Instant Messaging and Performance among Financial Traders}

\author{Serguei Saavedra\affil{1}{Northwestern Institute on Complex Systems, Northwestern University, Evanston, Illinois, USA, 60208}\affil{2}{Kellogg School of Management, Northwestern University, Evanston, Illinois, USA, 60208}, Kathleen Hagerty\affil{2}{Kellogg School of Management, Northwestern University, Evanston, Illinois, USA, 60208}, \and Brian Uzzi\affil{1}{Northwestern Institute on Complex Systems, Northwestern University, Evanston, Illinois, USA, 60208}\affil{2}{Kellogg School of Management, Northwestern University, Evanston, Illinois, USA, 60208} \thanks{To whom correspondence should be addressed. E-mail: uzzi@northwestern.edu}
}

\contributor{Submitted to Proceedings of the National Academy of Sciences of the United States of America}

\maketitle

\begin{article}
\begin{abstract}
Successful animal systems often manage risk through synchronous behavior that spontaneously arises without leadership. In critical human systems facing risk, such as financial markets or military operations, our understanding of the benefits associated to synchronicity is nascent but promising. Building on previous work illuminating commonalities between ecological and human systems, we compare the activity patterns of individual financial traders with the simultaneous activity of other traders---an individual and spontaneous characteristic we call synchronous trading. Additionally, we examine the association of synchronous trading with individual performance and communication patterns. Analyzing empirical data on day traders' second-to-second trading and instant messaging, we find that the higher the traders' synchronous trading, the less likely they lose money at the end of the day. We also find that the daily instant messaging patterns of traders are closely associated with their level of synchronous trading. This suggests that synchronicity and vanguard technology may help cope with risky decisions in complex systems and furnish new prospects for achieving collective and individual goals. 
\end{abstract}


\dropcap{S}ynchronous behavior has been found to enhance individual and group performance across a variety of domains even though the individuals might make no conscious effort to coordinate their behavior \cite{Strogatz03,Vicente,Arenas,Schneidaman,Sumpter,Couzin07}. Similarly, in systems of collaboration and competition, synchronous behavior can elude simple associations with individual benefits \cite{Sumpter,Ermentrout,Greenfield,Cooley,Cole}. Cicadas that chirp simultaneously with others find the best balance between risk and reward \cite{Cooley}. Cicadas that chirp in advance or in delay of the full chorus relish the best chances of finding a mate but may suffer the greatest risk of being spotted by a predator \cite{Ermentrout,Greenfield}. Congruently timed humans' actions have been surmised to provide potential benefits revealed by the meaningful coincidence of synchronous behavior \cite{Strogatz03,Merton,Galton,Neda,Page,Eckmann}. For instance, it has been found that simultaneous discoveries, or the times when multiple individuals arrive at a similar conclusion simultaneously, is collective evidence that a solution is valid \cite{Merton,Galton,Page}.
 
In this paper, we studied the association between individual performance and the simultaneous activity patterns followed by independent decision makers under risk. These conditions exist in many high-frequency decision contexts but are uniquely well documented in financial systems \cite{May}, where continuous change in information creates recurring uncertainty about when to trade and the second-to-second actions of financial traders are recorded \cite{Whaley,Moro}. Reducing the risk of losing money is the essence of trading \cite{Tetlock,Bouchaud,Brock,Whaley}. Over time the risks of trading can decrease as information is disambiguated. However, as this happens, the increasing certainty of information is incorporated into low-return prices. Thus, racing to be the first to discover the right time to trade is the critical problem to be solved \cite{Bouchaud,Brock}. By analogy, this optimal timing may represent the mating sweet spot observed for cicadas. Chirp in the sweet-spot and the chance of mating/returns is relatively high and predation/losses is relatively low. This suggests that as separate traders disambiguate their local view of news, they can spontaneously and simultaneously react as a group, without intention to coordinate, producing a synchronous behavior that might reveal the right time to trade in the market. Here, we tested whether traders' performance is relatively better when trading simultaneously with other traders---an individual and spontaneous characteristic we call synchronous trading.

Additionally, traders need to assess whether information is positive or negative for a stock, the potential magnitude of the information's impact, and the degree to which the information is already reflected in the price \cite{Bouchaud,Back}. Social corroboration is key to making these assessments \cite{Tetlock,Antweiller,Hong}. It reduces cognitive overload and ambiguity when diverse views converge on a common interpretation \cite{Page,Galton} and typically takes place among persons tied through social network relations \cite{Zhao}. We tested whether daily instant messaging patterns of traders are associated with the rise of synchronous trading. We believe our results can have broad implications for understanding fast collective action solutions to decision making under uncertainty. 

\section{Empirical Setting}

We observed all the second-to-second trades and instant messages of all 66 stock day traders in a typical trading firm from 9/07-02/09 (see Material and Methods). These day traders traded only stocks and made $>1$ million trades. $98.8\%$ of their trades are live, non-computerized trades (computerized trades were omitted from the analysis). Day traders typically do not hold stocks for more than a day. They typically enter and fully exit all their positions daily, which creates a standardized measure of performance: whether the trader made or lost money at the end of each day.  On average, these traders make money on just $55\%$ of their trades.

Despite sitting in the same firm, these traders generally trade independently rather than in teams. This is because they typically trade different stocks. One trader might trade high tech, one trader health care, another trader autos, and so on. Trading different stocks helps them diversify the firm's holdings, exploit their specialized sector knowledge, and avoid trading against each other. This means that traders have little incentive to simply mimic each other's trades or trading behavior. Nevertheless, despite trading different stocks, traders do process common market information. Common information includes Federal Reserve announcements, new job figures, housing market change, speculations about bankruptcies, or other global socio-economic data that traders attempt to disambiguate by exchanging information with others as they endeavor to discover the right time to trade \cite{Tetlock,Whaley}. A key form of information exchange here, and increasingly in other human complex systems, is instant messaging \cite{Leskovec}. Instant messages among traders and their network are based on elective relationships.  Each trader has the autonomy to communicate with persons of their choice. Hence, they are hierarchical ties in which orders are dictated from managers to subordinates. Within the communications, information is both professional and personal.  Typical information includes interpretations of market news, expectations of where the market might be moving, rumors, and forms of personal information commonly exchanged among business friends \cite{Centina}. 

Our extensive field research and interviews with traders at the firm confirmed that the content of these traders' messages included information consistent with the content found in other research \cite{Centina}. All the traders in the firm exchanged instant messages throughout the day with their network. Instant messages were sent and received from their terminals or mobile devices. By federal law all instant messages tied to trading go through the firm's capture system.  The importance of instant messages to these traders is instantiated by the intensity its use. We analyzed the full population of $>2$ million instant messages that our traders exchanged with their network of contacts in the industry.  

\section{Results}

\subsection{Synchronous trading}

To measure the synchronous trading of each trader with other traders, we defined a measure that quantifies the extent to which an individual's specific selection of time to trade is the same as the selection of other traders. To compare the synchronous trading among all traders across our observation period, we quantified the degree to which the number of traders $T_{ij}$ trading within the same time windows as trader $i$ in day $j$ compares to the same value when randomizing just the trades of trader $i$ (Fig. 1). Specifically, this randomization ensures normalizing individual activity, while keeping the trading structure and information heterogeneities of each specific trading day constant (e.g., number of traders, total number and timing of interactions, and number of interactions per trader). Mathematically, we defined synchronous trading as $s_{ij} = (T_{ij}- \langle T_{ij}^*\rangle)/\sigma_{T_{ij}^*}\,$, where $T_{ij}$ is the observed number of simultaneous traders and $\langle T_{ij}^*\rangle$ and $\sigma_{T_{ij}^*}$ are the average and standard deviation of simultaneous traders across an ensemble of random replicates within which the trades of trader $i$ were randomly shuffled. The greater the degree of a trader $i$'s daily $s_{ij}$, the greater her synchronous trading can be, and vice versa. Additionally, we defined the advanced trading $s_{ij}^{-1}$ and delayed trading $s_{ij}^{+1}$ in a similar fashion as the synchronous trading at zero-time lag $s_{ij}$, but quantified the number of traders trading one window late and one window early, respectively. 

We examined multiple time windows and reported the full analysis for 1-second windows. This window size was chosen for the main results for several reasons. First, the 1-second level of resolution comports well with the frenetic information environment and fast reaction time dynamics of modern markets \cite{Moro}. The time scale in which information heterogeneities exist has increasingly shortened with the growth of computerized trading, which now accounts for between $30\%$ to $60\%$ of the trading volume on financial exchanges \cite{Moro,Bouchaud}.  In computerized trading, preset algorithms trade very large volumes of shares in hundreds of a second, which means that traders must react to market opportunities that appear and disappear on a second-by-second scale. Also consistent with the view that information moves at high speeds in modern markets and that traders react at that level, we found that the traders in this firm do display a propensity to trade on 1-second time scales. The average empirical interval of consecutive 1-second trades is $1.01$ with a standard deviation of $0.14$. The maximum interval was $9$ and it occurred only twice in the data. Similarly, reaction time research has found that human reaction occurs in less than 1-second time frames \cite{Steven,Lo}. Second, we chose 1-second intervals for synchronous trading because it is the finest, most conservative time resolution in our data. Larger than 1-second intervals require a priori knowledge to find the appropriate balance between a window large enough to encapsulate changes in slow, non-computerized information heterogeneities and yet not so large that unrelated activities appear synchronized because they occur in a large window.  Working empirically to estimate this balance, we tested larger than 1-second intervals. We found that our results exist for intervals up to 15 seconds.  This window size seems to be a realistic limit for slower moving types of information and suggests that synchronicity is associated with individual thresholds that range across different information heterogeneities in this complex system \cite{Lo,Kiani}.

Our examination of the existence of synchronous trading revealed three interesting findings. First, Figure 2A pools all of our data and shows the probability density of synchronous trading at zero-time lag $s_{ij}$, advanced trading $s_{ij}^{-1}$ and delayed trading $s_{ij}^{+1}$. We observed that synchronous trading is significantly different ($p<10^{-3}$ using Kolmogorov-Smirnov test) to advanced and delayed trading. Values greater than $s_{ij}>10$ comprise $0.03\%$ of the entire data and are due to two traders, who could have a better access to information or better reaction times\cite{Lo,Kiani}. We conservatively omitted these outliers from our statistical analyses and found that our results did not change, confirming that synchronous trading is a special characteristic of collective behavior \cite{Schneidaman,Vicente,Greenfield}. This also suggests that timing is a key factor driving the decision of traders, and reminds us about the high-frequency changes in the market \cite{Moro}.

Second, Figure 2B shows that the average synchronous trading $\langle s_{ij}\rangle$ increases with the market's daily uncertainty ($p<10^{-4}$ using Markov randomizations), as given by the standard market volatility index, the VIX \cite{Whaley}. This finding supports the idea that collective behavior is associated with uncertainty as in the case of biological systems \cite{Couzin07,Torney}. The greater the level of uncertainty faced by individuals, the more likely is a collective behavior such as schooling or flocking to arise. These findings suggest a parallel in human systems. As the level of uncertainty in the market increases, the more likely is synchronous behavior to occur. Under high-uncertainty days, the average synchronous trading can increase to $\langle s_{ij}\rangle \approx 2$, i.e. the average synchronous trading of all traders is almost two standard deviations higher than the expected by chance.

Third, we found evidence that synchronous trading does not appear to be due to coordination. Unlike coordinated behavior, where pairs or sets of actors consistently align their behavior, synchronous activity commonly displays the opposite pattern \cite{Schneidaman}. We found that $98\%$ of all pairwise correlations between activity patterns of two different traders are non-significant $p>0.15$. This is probably because no two individuals consistently follow the same strategic behavior and the same two actors are not always correct in their assessments about the market. Similarly, if synchronous trading was driven by coordination, we should observe simultaneous trades of predominantly the same stock \cite{Bouchaud,Brock}. However, we find that 96\% of our simultaneous trades are of different stocks. Moreover, the trades are of different types: 60\% of the simultaneous trades involve both buying and selling activities. 
 
\subsection{Individual performance}

Individual daily performance $p_{ij}$ can be assessed by whether the trader $i$ loses or makes money at the end of the day $j$. Moreover, because the amount of money made by a trader at day's end depends on various factors, such as market volatility, number of stocks traded, and size of trades, a simple binary outcome variable appropriately standardizes their performance by considering whether the trader lost ($\$<0$) or made money ($\$>0$). This is coded as $p_{ij}=0$ and $p_{ij}=1$, respectively. 

We quantified the relationship between a trader's synchronous trading $s_{ij}$ and performance $p_{ij}$ with a logistic regression of the form ${\rm logit}(p_{ij}) = \beta_0 + \beta_1 s_{ij}$ (see Materials and Methods). We found that synchronous trading $s_{ij}$ was significantly ($p<10^{-3}$) and positively associated with a trader's performance (Fig. 3). Using the same logistic analysis, we compared advanced and delayed trading with end-of-day performance $p_{ij}$. The results indicate that both advanced and delayed trading are statistically unrelated ($p>0.15$) to end-of-day performance. This reveals that synchronous trading, though arising without apparent coordination, indicates a uniquely beneficial time to trade that neither advanced nor delayed trading can reveal.

\subsection{Instant messaging patterns}

An important proposed contribution of our work is to identify those factors associated with the level of collective human behavior. In biological systems, local communication channels have been identified as a correlate of the rise of synchronous behavior \cite{Strogatz03,Arenas,Cole,Strogatz05,Couzin07}. Following this line of reasoning, we found distinctive associations between traders' instant messaging patterns and synchronous trading. First, we found that instant messaging volume is associated with trading volume throughout the day, suggesting a close connection between the two. Pooling all traders' IM and trade activity over our observation period, Figure 4A shows that IMs have a significant correlation ($p<10^{-10}$) with trades over the day. On average, IM and trade activity rise rapidly after the 9:30am opening bell, peak at 10am, decline at lunch time, uptick again from 1-3 pm, and finally decline precipitously at the 4 pm closing bell.

Second, research has shown that collective synchronous activity can arise when a coupling mechanism delays or pauses the timing of individual activities \cite{Strogatz03,Arenas,Vicente,Hunt,Greenfield,Hunt2}. For example, cicadas have been found to have an internal clock that stimulates chirping. This clock would induce a cicada to regularly chirp whether or not it was exposed to the chirps of other cicadas.  Synchrnous chirping arises because the internal and individual chirp activity is delayed by the chirp of another cicada, coupling the timing of the internal chirp with the collective chirps of other cicadas \cite{Ermentrout}. To determine if instant messages can play a coupling role, we observed whether they were associated with the rise of synchronous trading. As noted above, instant messages play the important function of transferring information that helps traders disambiguate market information, and, because a trader cannot trade and IM at the same time, instant messaging necessarily can delay his or her trades. This would suggest that if we were to compare the observed intensity of synchronous trading of a trader relative to the synchronous trading expected by chance, as we did above, but this time we randomized the trader's trades across all 1-second time windows of the day except those 1-second time windows where there was an instant message, we would expect the intensity of synchronous trading to increase in the presence of non-random instant messages. To quantify the IM-trade coupling, we compared the degree to which synchronous trading changes when randomizing a single trader's trades in any second in which an IM was not sent or received. Methodologically, we quantified the IM-trade coupling by $\theta_{ij}=|s_{ij}-\hat{s_{ij}}|$, where $\hat{s_{ij}}$  is the synchronous trading calculated by randomly shuffling over the seconds with no IM activity.

Consistent with our conjectures, we found a positive and significant association ($p<10^{-10}$ using Markov randomizations) between the IM-trade coupling $\theta_{ij}$ and the synchronous trading $s_{ij}$ of each trader (Fig. 4B, Materials and Methods). This reveals that the instant messaging patterns of traders is associated to their trades such that the observed level of synchronous trading increases as the communication pattern is increasingly different from what would be expected by chance. The more non-random the instant messaging pattern, the greater the synchronous trading. This suggests that the local communication patterns of individuals have an important association with the rise of simultaneous activity, which in turn is associated to their performance.  If one assumes that IMs are used to corroborate the meaning of the market throughout the day, then our findings suggest that the increasingly structured communication strategically aims to help each individual trader make a decision about when to trade.

\section{Discussion}

Synchronicity is a pervasive and mysterious drive in nature \cite{Strogatz03}. In animal, biological, and physical systems synchronicity reduces uncertainty, as when school of anchovies evade predation, neurons co-fire to process complex information, or perturbations reduce noise in physical systems. Synchronicity also apparently arises from local interactions without the aid of centralized leadership. This suggests that while more research needs to be done on synchronicity's functional role in complex human systems, it may furnish a functional alternative to leadership in rich informational environments.  
	
We examined the association between synchronicity and performance in a complex system where performance increases with uncertainty reduction. Examining a typical proprietary trading firm wherein the traders individually race to be the first to disambiguate a constant stream of uncertain market information in an effort to make profitable trades, we found that when a stock trader in that firm trades at the same time as other traders in the firm, his or her financial performance is significantly increased. We also found a coupling mechanism; we found that traders' instant message communication patterns are positively associated with the rise of sync. Building on synchronicity principles found in other complex systems, we speculate that the mechanisms underlying these empirical patterns involve rapid information aggregation through instant messaging networks. Because separate traders in the firm have different instant message contacts in the market, each trader samples, separate local inferences about the eventual meaning of market information. When these diverse points of view converge, the traders trade in synchronicity such that the synchronous timing of trades reflects a point of crowd wisdom despite no conscious intention to do so on the part of any individual trader. These mechanisms suggest that synchronicity in human systems reflect some of the same principles found in animal systems, namely that synchronicity appears to arise with attention to local information rather than centralized leadership. In the human system we examine, and in human systems where quick response behavior is likely a mix of being reactive and thoughtful about the information presented, we also purport that the rapid aggregation of local information from diverse points of view plays a role in the performance benefits of the synchronicity we observe. If one assumes that each of the traders have their own expertise, training, and assumptions that go into deriving inferences from the market information they process, it suggests that actions consistent with the corroboration of diverse viewpoints are likely to be a better approximation of the true meaning of information than singular or myopic points of view.

The powerful information processing capacities of humans in complex systems may furnish unique opportunities to apply the ideas developed here about human synchronicity to other contexts. We would speculate that in many increasingly rich information environments there are benefits to understanding synchronicity. For example, currently in the domain of intelligence and national security, many security officers face a frenetic pace of information not unlike the traders we studied. They too receive constant feeds of information---videos, text, voice, blogs, RSS newsfeeds, and tweets---and are in constant communication with their own instant message network throughout the day. Moreover, like traders they are also racing to disambiguate news. Disambiguating information quickly means a potential pre-empt of an attack whereas advanced or delayed disambiguation can mean ``jumping the gun" or waiting too long respectively \cite{Kiani}. Disease control agencies around the world all monitor large amounts of time sensitive data in attempts to identify possible outbreaks. In both situations, and more generally, in situations where information overloads might overwhelm individual decision making ability and information is time sensitive, creating systems that can capture moments of synchronicity may help identify whether an action is functional or not.

Using observational data to describe this phenomenon provided a rich mix of real data but we are unable to completely test these mechanisms.  Future work might devise experiments, perhaps in one of the mock trading labs now in existence in universities, by manipulating the content and rate of change of market news, providing access vs. no access to instant message communication networks, and changing the information sampled from the instant message networks from myopic to diverse. Future research might also begin to examine the potential dysfunctions of synchronicity in human complex systems. Under what conditions does collective genius turn into mob madness?  Another direction for future research is to explore the differences between synchronicity and other collective behavior mechanisms.

\begin{materials}
\section{Data} We observed all the 66 day traders at an anonymous trading firm from 9/26/2007-02/20/2009.  Day traders keep short-term positions and do not hold inventories of stocks; they enter and exit positions each day, normally between 9.30am-4:00pm.  Our traders are ``point-and-clickers"-- they make trades in real-time in 98\% of the time (the 1.2\% of the trades done algorithmically were omitted and did not affect the results).  40-70\% of the trading on NYSE is point-and-click.  We observed these traders trading approximately 4500 different stocks over various exchanges, which suggests that they sample a large part of the market.  As in most trading firms, traders do not trade everyday of every week for various reasons.  Similarly, in this firm, no more than 22 traders were at their desks on any one day  We analyzed all the $>1$ million intraday stock trades of these day traders and their $>2$ million instant messages exchanged across their networks. The performance data was calculated by the firm using standard industry metrics. Traders cannot trade via IMs.\\
\section{Logistic regression} To check the robustness of the association between synchronous trading and individual performance to other potential influences such as number of trades and daily uncertainty, we performed the same analysis with terms for number of trades, an interaction term for number of trades and $s_{ij}$, and we also controlled for market volatility (i.e. VIX). The extended model has the form ${\rm logit}(p_{ij}) = \alpha + \beta s_{ij} + \gamma k_{ij} + \delta s_{ij} k_{ij} + \epsilon v_j$. Additionally, to control for unobservable factors of each particular trader, we used fixed effects (dummy variables for each trader) in the logistic regression. Under all circumstances, synchronous trading was significantly and positively associated with individual performance.\\
\section{Association between synchronous trading and IM-trade coupling} To check the robustness of the association between IM-trade coupling and synchronous trading to other potential influences such as number of IMs and daily uncertainty, we performed a linear regression with terms for number of IMs, an interaction term for number of IMs and $\theta_{ij}$, and we also controlled for market volatility (i.e. VIX). The extended model has the form $s_{ij} = \alpha + \beta s_{ij} + \gamma \theta_{ij} + \delta s_{ij} \theta_{ij} + \epsilon v_j$. Additionally, to control for unobservable factors of each particular trader, we used fixed effects (dummy variables for each trader) in the regression. Under all circumstances, IM-trade coupling was significantly and positively associated with synchronous trading.
\end{materials}

\begin{acknowledgments}
We would like to thank Alex Arenas, Jordi Bascompte, Jordi Duch, William Kath, Tae-Hyun Kim, Eduardo L\'{o}pez, Hani Mahmassani, Dean Malmgren, Alejandro Morales Gallardo, Mason Porter, Mark Rivera, Daniel Stouffer, Felix Reed-Tsochas, Marta Sales-Pardo, Uri Wilensky, and the members of the NICO weekly seminar series for useful discussions that led to the improvement of this work. We also thank the Kellogg School of Management, Northwestern University, the Northwestern Institute on Complex Systems (NICO) for financial support. Research was also sponsored by the Army Research Laboratory and was accomplished under Cooperative Agreement Number W911NF-09-2-0053. The views and conclusions contained in this document are those of the authors and should not be interpreted as representing the official policies, either expressed or implied, of the Army Research Laboratory or the U.S. Government. The U.S. Government is authorized to reproduce and distribute reprints for Government purposes notwithstanding any copyright notation here on.
\end{acknowledgments}

\end{article}

\begin{figure*}
\centerline{\includegraphics*[width=3.5in]{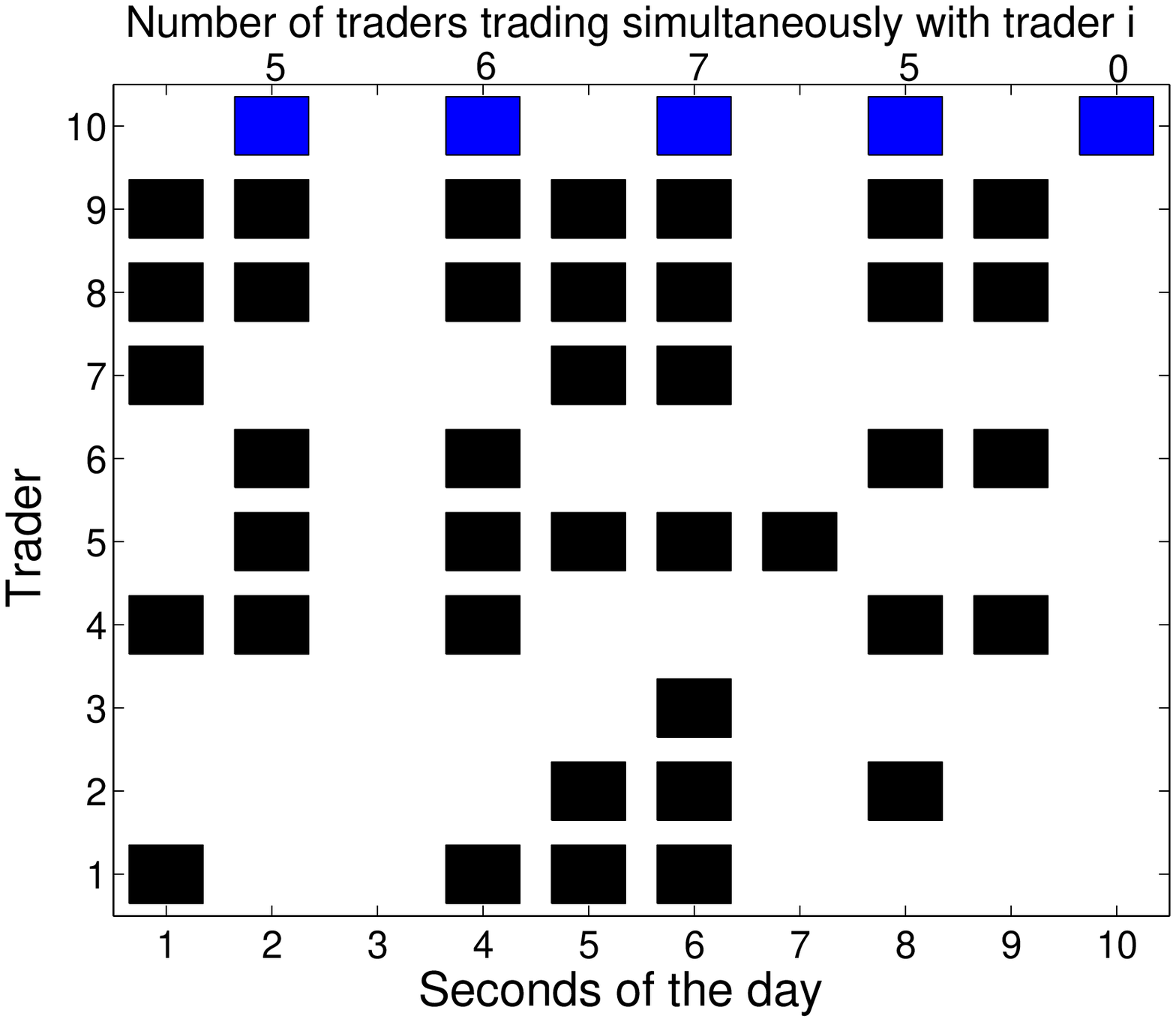}}
\caption{Calculating synchronous trading. The synchronous trading $s_{ij}$ of a trader $i$ in day $j$ (e.g., the trader whose trades are highlighted in blue) is defined as the degree to which the observed number of other traders trading within the same seconds as trader $i$ (top values) compares to the same value when randomizing just the trades of that particular trader. For advanced and delayed trading, we calculated the number of other traders trading one second late and one second early, respectively.}
\label{fig1}
\end{figure*}

\begin{figure*}
\centerline{\includegraphics*[width=7in]{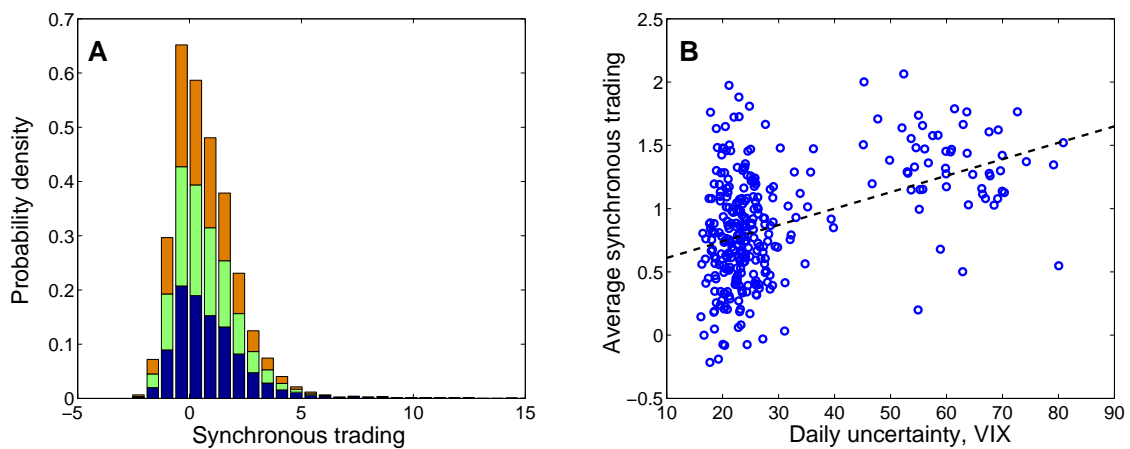}}
\caption{Synchronous trading and uncertainty. {\bf A} shows the probability density for synchronous trading $s_{ij}$ (bottom blue bars), advanced trading $s_{ij}^{-1}$ (middle green bars) and delayed trading $s_{ij}^{+1}$ (top orange bars) for all traders across the observation period. The bar size is the sum of the 3 probability values, and colors correspond to the relative contribution each distribution makes to the total sum.  We found that synchronous trading is significantly different to advanced and delayed trading ($p<10^{-3}$ using Kolmogorov-Smirnov test). {\bf B} shows the positive association ($p<10^{-4}$ using Markov randomizations) between the average synchronous trading $\langle s_{ij}\rangle$ and level of daily uncertainty in the market, as given by the market's standard volatility index (VIX) \cite{Whaley}. The dashed line depicts the relationship estimated via a linear regression.
}
\label{fig2}
\end{figure*}

\begin{figure*}
\centerline{\includegraphics*[width=7in]{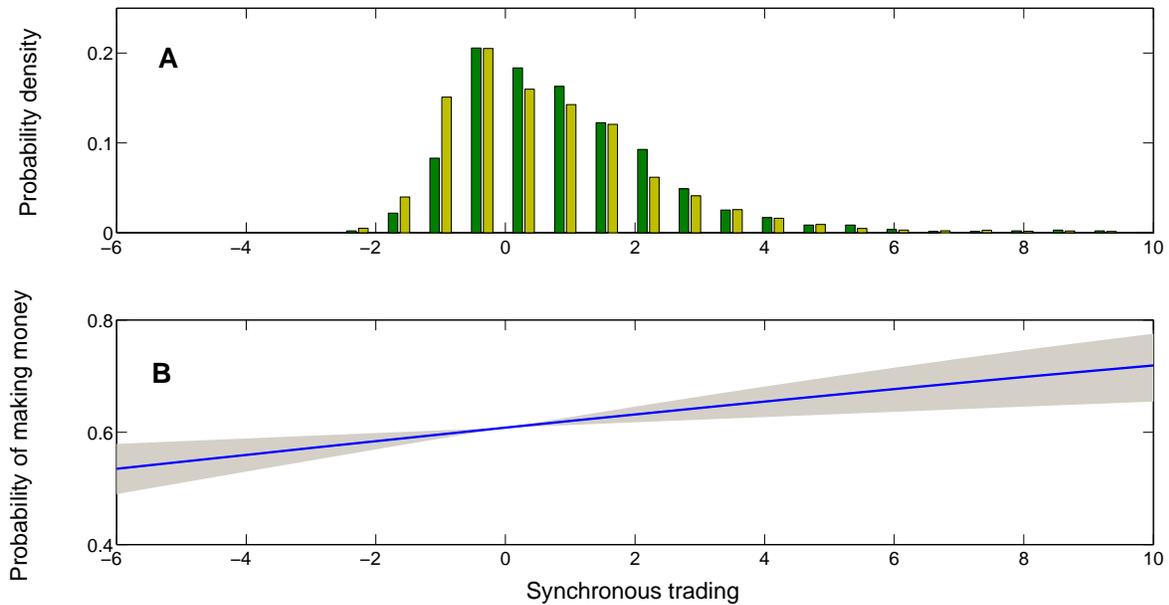}}
\caption{Individual performance. {\bf A} depicts the probability density of synchronous trading for traders that make money (left green bars), and for those that do not (right yellow bars). The two distributions are significantly different considering all values ($p=0.004$ using Kolmogorov-Smirnov test), within -2 and 2 exclusively ($p=0.046$ using Kolmogorov-Smirnov test) and outside -2 and 2 ($p=0.038$ using Kolmogorov-Smirnov test). {\bf B} shows the relationship between synchronous trading $s_{ij}$ and the probability of making money $p_{ij}$. The curve depicts the probability of performance (making money) estimated via a logistic regression (Methods). For \emph{any} trader under consideration, the probability of making money increases as the synchronous trading increases. The gray region corresponds to the $95\%$ confidence interval.
}
\label{fig3}
\end{figure*}

\begin{figure*}
\centerline{\includegraphics*[width=7in]{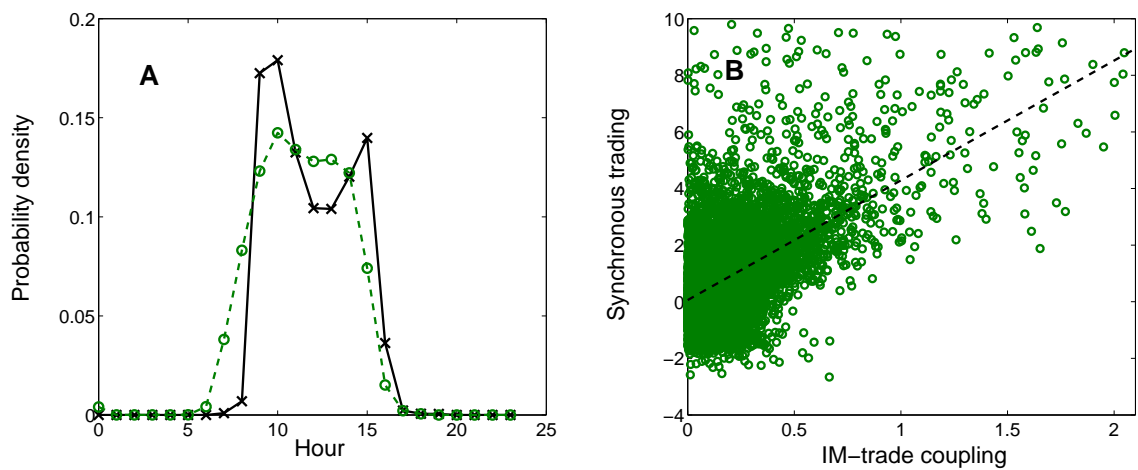}}
\caption{Association between IMs and trading. Panel {\bf A} presents the probability density of observing any trade (black line) and IM (green dashed line) in each hour on average across the observation period.  Approximately 95\% of trades and IMs are done between 9.30am and 4pm, which correspond to the main operation hours of NYSE. Panel {\bf B} shows the empirical relationship ($p<10^{-10}$ using Markov randomizations) between the IM-trade coupling $\theta_{ij}$ and synchronous trading $s_{ij}$ for all traders across the observation period. The dashed line depicts the association estimated via a linear regression (Methods).
}
\label{fig4}
\end{figure*}

\end{document}